\RequirePackage{ifpdf}
\RequirePackage{color}
\documentclass{PoS}
\usepackage[authoryear]{natbib}
\bibpunct{(}{)}{;}{a}{}{,}
\usepackage{graphicx}
\usepackage{verbatim}
\usepackage{url}
\usepackage{mathrsfs}
\usepackage{amssymb}
\usepackage{amsmath}
\usepackage{bm}
\usepackage{color}
\usepackage[table]{xcolor}
\usepackage{multirow}
\usepackage{caption}
\usepackage{subfig}
\usepackage{afterpage}
\usepackage{floatrow}
\newcommand{\etal}{{\it et al}}

\title{Morphological classification of radio sources for galaxy evolution and cosmology with SKA-MID}

\ShortTitle{Morphological classification of radio sources  for galaxy evolution and cosmology with SKA-MID}

\author{S. Makhathini$^1$, O. M. Smirnov$^{12}$, M. J. Jarvis$^{34}$ and I. Heywood$^{51}$\\
  \llap{$^1$} Department of Physics \& Electronics, Rhodes University, P.O. Box 94, Grahamstown 6140, South Africa\\ 
  \llap{$^2$} SKA South Africa, 3rd Floor, The Park,
Park Road, Pinelands, 7405, South Africa \\
  \llap{$^3$} Astrophysics, Department of Physics, University of Oxford, Oxford OX1 3RH, UK\\ 
  \llap{$^4$} University of the Western Cape, ZA-7535 Cape Town, South Africa\\
  \llap{$^5$} CSIRO Astronomy \& Space Science, P.O. Box 76, Epping, NSW 1710, Australia\\
      E-mail: \email{sphemakh@gmail.com}, \email{o.smirnov@ru.ac.za}, \email{matt.jarvis@astro.ox.ac.uk}, \email{ian.heywood@csiro.au}}

\abstract{Morphologically classifying radio sources in continuum images with the SKA has the potential to address some
of the key questions in cosmology and galaxy evolution. In particular, we may use different classes of radio sources as
independent tracers of the dark-matter density field, and thus overcome cosmic variance in measuring large-scale
structure, while on the galaxy evolution side we could measure the mechanical feedback from FRII and FRI jets. This work
makes use of a \texttt{MeqTrees}-based simulations framework to forecast the ability of the SKA to recover true source
morphologies at high redshifts. A suite of high resolution images containing realistic continuum source distributions
with different morphologies (FRI, FRII, starburst galaxies) is fed through an SKA Phase 1 simulator, then analysed to
determine the sensitivity limits at which the morphologies can still be distinguished. We also explore how changing the
antenna distribution affects these results.}

\FullConference{
Advancing Astrophysics with the Square Kilometre Array\\
June 8-13, 2014\\
Giardini Naxos, Italy}

\begin{document}

\section{Introduction}
The capabilities of the Square Kilometre Array (SKA) allow it to conduct experiments in areas of fundamental physics and
cosmology with accuracies that are unprecedented in the field of radio 
astronomy~\citep[][this volume]{cordes2004,rawlings2004,kramer2004}.

One key experiment that forthcoming radio continuum surveys will be able to perform involves the investigation of large 
scale structure formation in the Universe. The inhomogeneous distribution of matter in the Universe is thought to
be seeded by random perturbations in the density field imprinted shortly after cosmic inflation. The magnitude of
these primordial fluctuations are typically investigated by measuring the angular two-point correlation 
functions~\citep[2PCF; e.g.][]{peebles1980}\footnote{The Fourier inversion of which is the matter power spectrum} of 
both galaxy catalogues~\citep[e.g.][]{wang2013} 
and the temperature fluctuations in the Cosmic Microwave Background~\citep[e.g.][]{spergel2003}. Evidence for 
non-Gaussianity in the density field
has implications for inflationary models, and can be investigated by determining the bispectrum\footnote{The
Fourier inversion of which is the three point correlation function}, specifically the non-linearity function 
$f_{NL}$~\citep[2.7~$\pm$~5.8;][]{ade2013}.

One of the best methods for constraining $f_{NL}$ involves the use of galaxy surveys to trace the dark matter
distribution at more recent cosmic epochs. It is necessary for these
surveys to cover very large areas in order trace the large-scale
power. The SKA will have the sensitivity to detect a large number of very faint radio sources (which are
typically all at cosmological distances) over large areas of the sky. However one advantage the radio wavelength has 
over optical surveys is the possibility of morphologically distinguishing different source populations.

Three such populations of sources are the two Fanaroff-Riley~\citep[FR;][]{fanaroff1974} classes of jet-producing 
active galaxies, and regular star forming disks exhibiting synchrotron emission at radio wavelengths, and possibly 
hosting a weak active nucleus. Each of these sources has a distinct morphological appearance, and coupled with the 
correlation between the source type and the halo mass in which it resides, the uncertainty on $f_{NL}$ could be 
conservatively halved by a plausible SKA continuum survey~\citep{ferramacho2014}.
Characterising radio source morphology is also critical for the vast majority of galaxy evolution and AGN-related
science. It is now clear that feedback from AGN is a critical mechanism in the evolution of massive galaxies over all
redshifts. One of the few methods of identifying the sources responsible for the hot-mode, or radio-mode feedback is
through radio continuum observations, as indications of such activity at other wavelengths are not widely 
observed~\citep{BestHeckman2012}. Furthermore, radio morphologies can help distinguish AGN from 
star-forming galaxies, which will dominate the source counts at the flux-densities that the SKA will 
reach~\citep[][this volume]{muxlow2005,McAlpine2015}
This is crucial if we are to use the SKA to provide a robust, obscuration free, method of 
determining the evolution of the star-formation rate density in the
Universe~\citep[][this volume]{muxlow2005}. However, one of the 
most crucial elements of good morphological information is to separate
star formation and AGN activity~\citep[see e.g.][this volume]{McAlpine2015} in the same 
galaxy, thus allowing us to attempt to decouple the bolometric output from these two processes.
Key to this experiment is the radio survey having sufficient spatial resolution and imaging fidelity in order to
faithfully reproduce the source morphologies in the synthesised image.
In this chapter we consider a limited class of radio sources in an attempt to forecast the ability of 
the SKA1-MID array to detect and morphologically distinguish between these sources. We concentrate on instrumental 
limitations to morphological classification (sensitivity, resolution, imaging fidelity); questions of the relative size 
of morphologically distinct source populations are outside the scope of this work. In other words, we attempt to 
answer the following question: if the sources are morphologically distinct in the radio, can we make this distinction 
with SKA1-MID, and how deeply?
. We use plausible SKA configurations and perform full imaging simulations of schematic representations of various 
source types. The redshift, signal to noise  and resolution limits on reliable source classifications are determined.

\section{Background on Layouts}\label{sec:layouts}
\begin{figure}[!ht]
 \begin{tabular}{cc}
\includegraphics[width=0.450000\textwidth,trim= 0 .05cm 0 0.05cm]{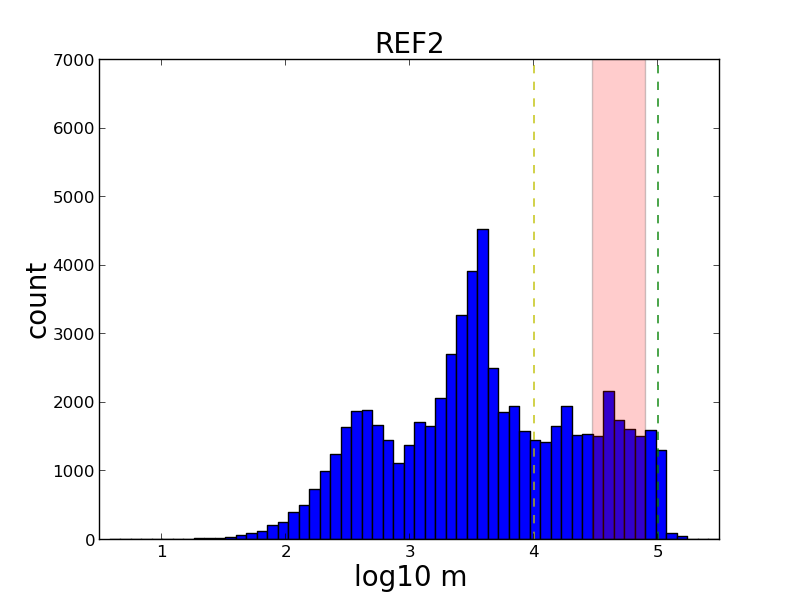} 
&\includegraphics[width=0.450000\textwidth,trim= 0 .05cm 0 0.05cm]{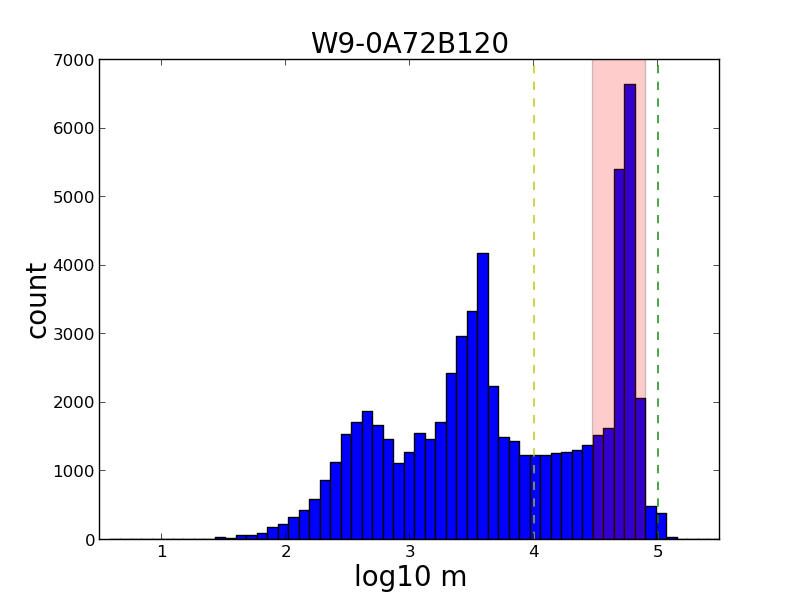} 
 \\\end{tabular}
 \caption{Baseline distribution with the uv-distance in log$_{10}$~km . Yellow and green dashed lines mark 10 and 120
kilometres respectively, and the pink strip represents baselines from 30-80~km.}\label{fig:hist}
\end{figure}

The general scientific requirements for SKA1~\citep{ska2014} suggests that (at least for SKA1-MID), an array with a 
maximum baseline of around 100~km is required. Therefore, we consider a layout with the shortest possible maximum 
baseline that does at least as  well as the ``second generation'' baseline design in the resolution range 0.4-1$''$ 
over 650, 800 and 1000 MHz while not significantly compromising the performance at the larger angular scales. Moreover, 
having a layout which performs just as well as (or better than) the baseline layout but which covers significantly less 
space translates to a reduction in trenching and data transport costs, which presents an opportunity to re-invest the 
funds elsewhere. The following SKA1-MID layouts are under consideration here:
\begin{description}
\item[{\bf REF2A100B173}] The ``Second-generation'' layout (254 dishes) produced by Robert Braun (September
2013)\footnote{We assume this to be the baseline layout.}. This layout has a maximum baseline of 173~km. In this 
chapter, we also refer to this layout as REF2.
\item[{\bf W$i$-$j$A$k$B$l$}] This is the REF2 layout with the core ``puffed up'' by 10\%,  with $i$ dishes moved  from
the outer core to the spiral arms and $j$ extra dishes added to the spiral arms. The spacing in the arms is then
optimised  to get more sensitivity on the longer ($>50$~km) baselines (See baseline distribution histograms in Figure
\ref{fig:hist}). Each spiral arm stretches out to~$k$ kilometres and the maximum baseline length is about $l$
kilometres.
\end{description}
\begin{figure}[!ht]
 \begin{tabular}{lll}
\includegraphics[width=0.300000\textwidth,trim= 0 .05cm 0 0.05cm]{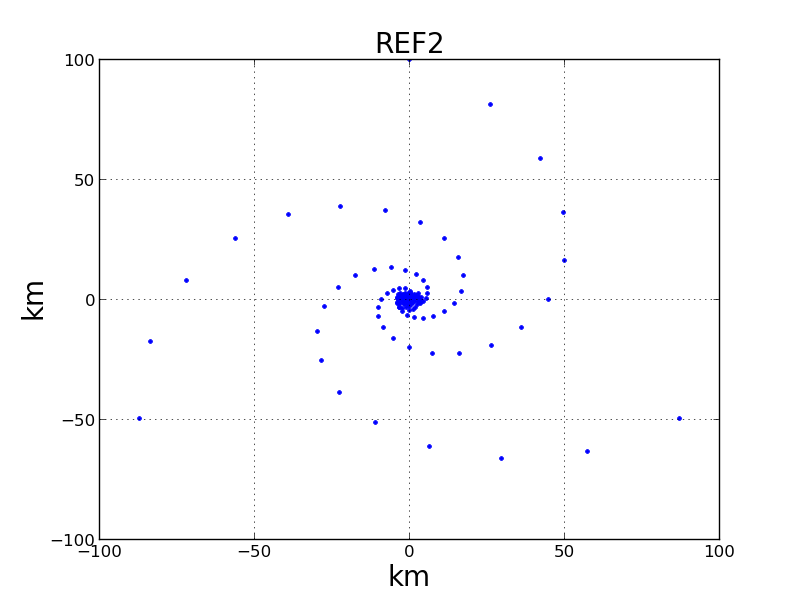} 
&\includegraphics[width=0.300000\textwidth,trim= 0 .05cm 0 0.05cm]{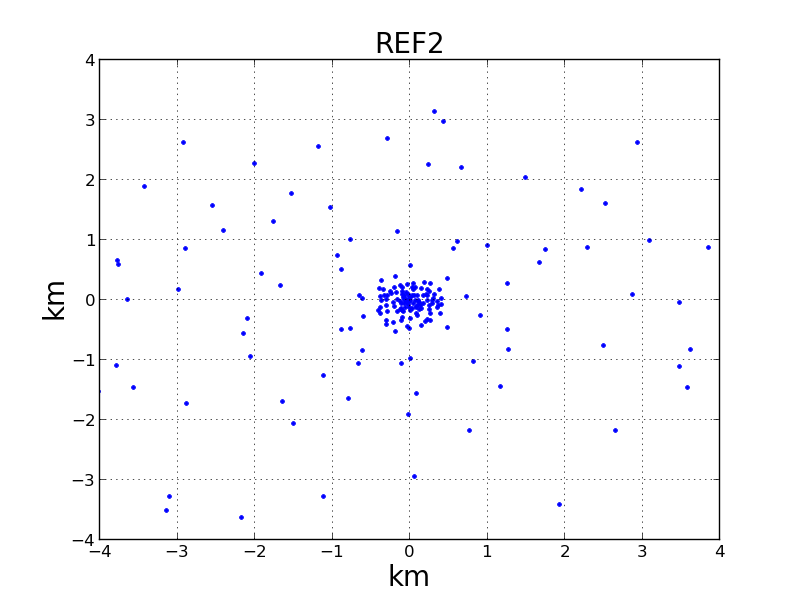} 
&\includegraphics[width=0.300000\textwidth,trim= 0 .05cm 0 0.05cm]{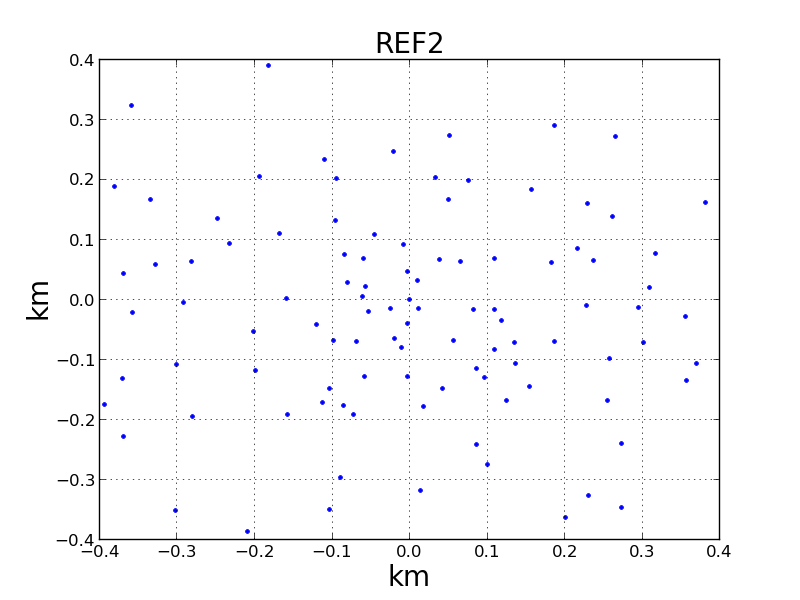} 
 \\ \hfill\includegraphics[width=0.300000\textwidth,trim= 0 .05cm 0 0.05cm]{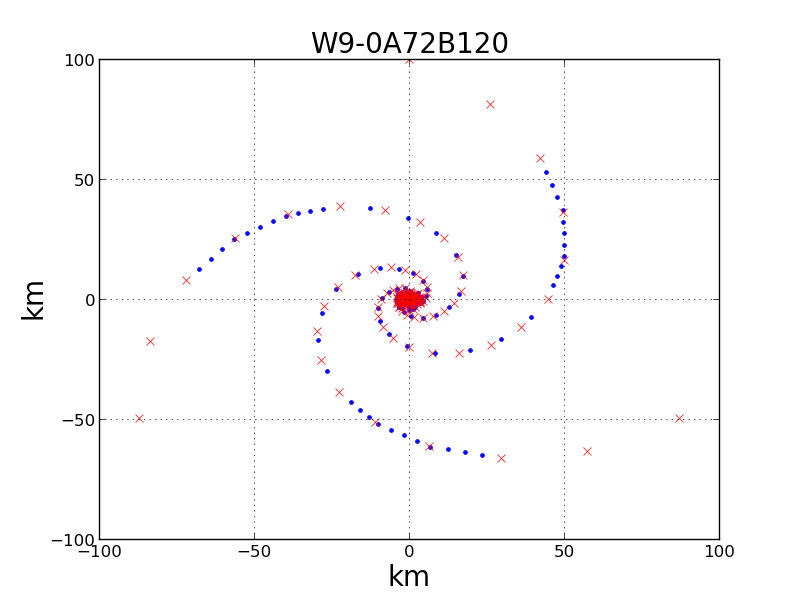} 
&\includegraphics[width=0.300000\textwidth,trim= 0 .05cm 0 0.05cm]{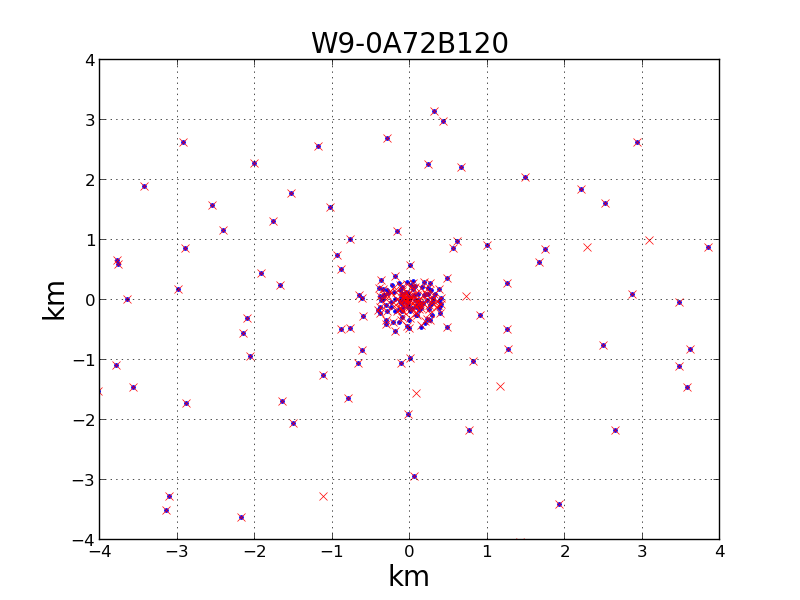} 
&\includegraphics[width=0.300000\textwidth,trim= 0 .05cm 0 0.05cm]{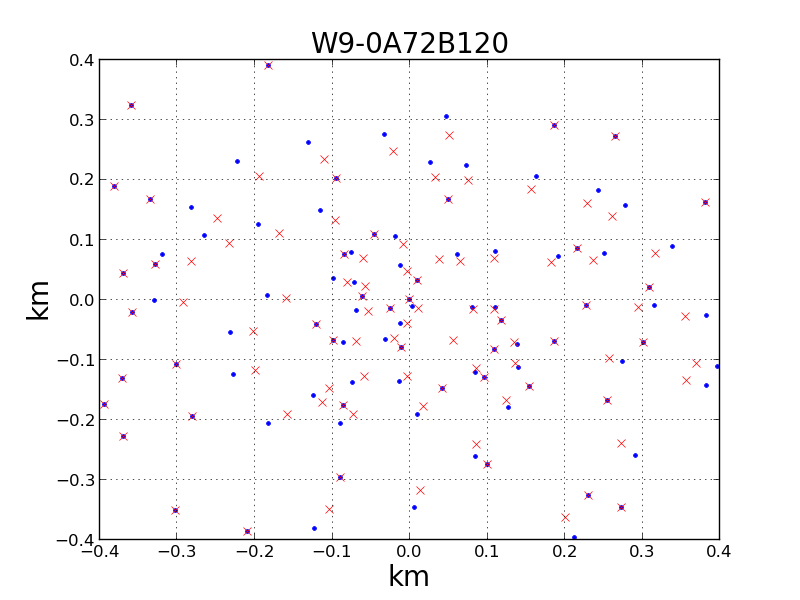} 
 \\ \hfill\end{tabular}
 \caption{Antenna layouts on scales of $\pm$100~km (left), $\pm$4~km (centre) and $\pm$400~m (right). The REF2 layout
 is plotted on the upper row, and as red crosses for direct comparison to the W9-0A7B120 configuration on the lower 
row.}\label{fig:lay}
\end{figure}

\begin{figure}[!ht]
 \begin{tabular}{cc}
\includegraphics[width=0.450000\textwidth,trim= 0 .05cm 0 0.05cm]{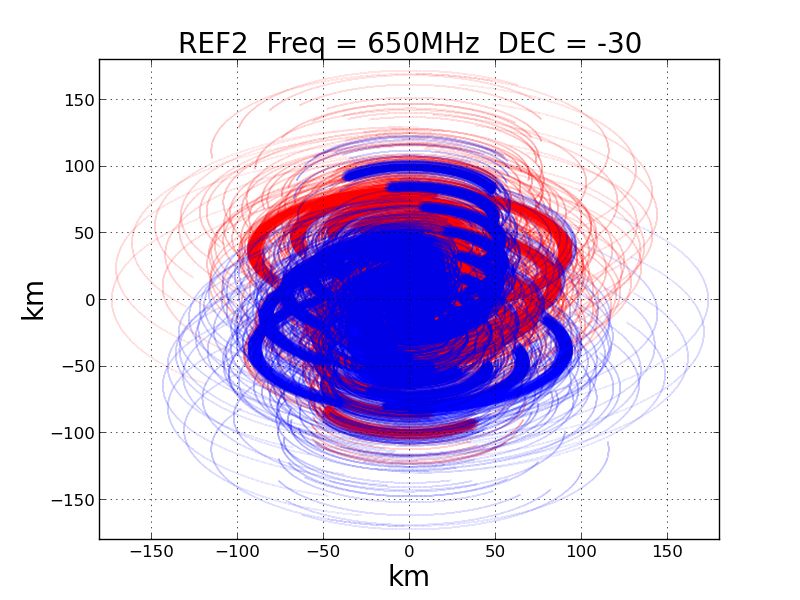} 
&\includegraphics[width=0.450000\textwidth,trim= 0 .05cm 0 0.05cm]{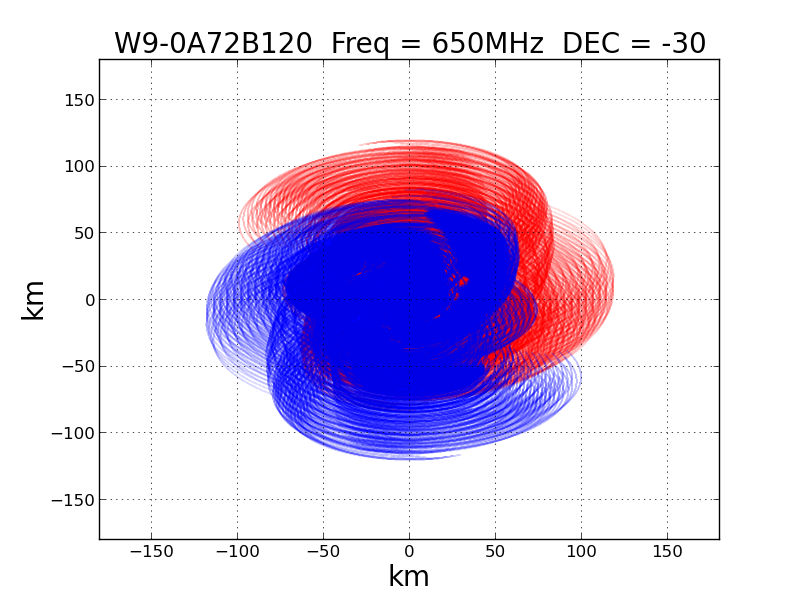} 
 \\\end{tabular}
 \caption{UV-Coverage for 8 hour tracks at 650~MHz (50~MHz wide channel) at declination -30 degrees for the different 
layouts. Blue indicates uv-points, red indicates conjugate uv-points.}\label{fig:uvcov}
\end{figure}

\begin{figure}[H]
 \includegraphics[width=0.7\textwidth]{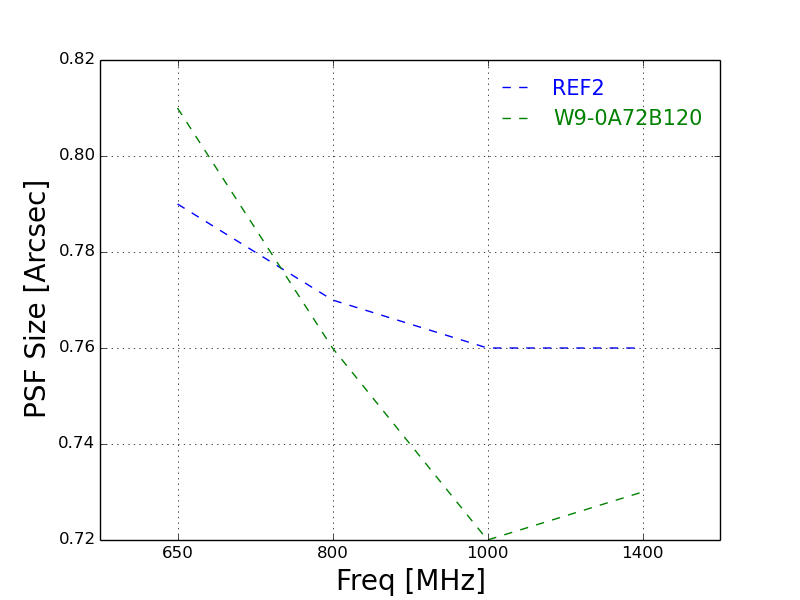}
 \caption{Uniformly weighted PSF sizes as a function of frequency for the REF2 (blue) and W9-0A72B120 (green) 
layouts.}\label{fig:psf_mean}
\end{figure}

The natural sensitivity at specific angular scales for an interferometer can be determined by making noise maps using 
only visibilities corresponding to those angular scales. Table \ref{tab:noise50-chap} shows the sensitivity on angular 
scales \{0.4-1, 1-2, 2-3, 3-4, 600-3600\} arcsec for the two layouts under consideration. It is evident from this table 
that the W9-0A72B120 layout (henceforth, referred to as W9) has more sensitivity at small angular scales, but we note 
that further optimisations are still possible -- even for different angular scales. In fact, the sophistication of 
simulation packages like {\tt MeqTrees}~\citep{noordam2010} coupled with new insights into fundamental limits on 
radio interferometric imaging and calibration~\citep{stefan2008} allow for the distribution of array elements to be 
optimised subject to a well defined set of science of goals. It is also worth noting that the optimisation can be 
further constrained by cost and engineering limitations. Figure \ref{fig:psf_mean} shows the PSF sizes (uniform 
weighting) for the two layouts under consideration. The figure shows that the 120km W9 layout has very similar 
resolving 
performance compared to the 173km REF2.

 \begin{table}[!htp]
 \tiny{
\subfloat{\begin{tabular}{|lccccc||ccccc||ccccc|} 
 \tabularnewline \cline{2-16} \multicolumn{1}{c}{ } & \multicolumn{5}{|c}{650MHz}  & \multicolumn{5}{c}{800MHz}  & 
\multicolumn{5}{c|}{1000MHz} \tabularnewline \cline{1-16} 
 resbin  &1 & 2 & 3 & 4 & 5 & 1 & 2 & 3 & 4 & 5 & 1 & 2 & 3 & 4 & 5 \tabularnewline \hline
SKA1REF2 & 1.00 \cellcolor{blue!60.00} & 1.00 \cellcolor{red!60.00} & 1.00 \cellcolor{green!18.00} & 1.00 
\cellcolor{orange!18.00} & 1.00 \cellcolor{purple!18.00} & 1.00 \cellcolor{blue!60.00} & 1.00 \cellcolor{red!18.00} & 
1.00 \cellcolor{green!18.00} & 1.00 \cellcolor{orange!18.00} & 1.00 \cellcolor{purple!18.00} & 1.00 
\cellcolor{blue!60.00} & 1.00 \cellcolor{red!18.00} & 1.00 \cellcolor{green!18.00} & 1.00 \cellcolor{orange!18.00} & 
1.00 \cellcolor{purple!18.00}\\ \hline 

SKA1W9-0A72B120 & 0.77 \cellcolor{blue!18.00} & 0.81 \cellcolor{red!18.00} & 1.12 \cellcolor{green!60.00} & 1.06 
\cellcolor{orange!60.00} & 1.00 \cellcolor{purple!18.00} & 0.72 \cellcolor{blue!18.00} & 1.05 \cellcolor{red!60.00} & 
1.06 \cellcolor{green!60.00} & 1.17 \cellcolor{orange!60.00} & 1.06 \cellcolor{purple!60.00} & 0.72 
\cellcolor{blue!18.00} & 1.05 \cellcolor{red!60.00} & 1.08 \cellcolor{green!60.00} & 1.26 \cellcolor{orange!60.00} & 
 1.05\cellcolor{purple!60.00}\tabularnewline \hline 
\end{tabular}}\hfil 

\caption{Relative (w.r.t REF2) RMS pixel noise for a 50MHz band after an 8hr synthesis with a 60s integration for 
the different layouts at different angular scales. These values are generated at 650, 800 and 1000 MHz, at angular 
scales \{0.4-1, 1-2, 2-3, 3-4, 600-3600\} arcsec and are labeled {\it resbin} \{1, 2, 3, 4, 5\} respectively. This is 
done for natural weighting at declination -30 degrees.}\label{tab:noise50-chap}}
 \end{table}

\section{The Experiment}\label{sec:exp}
Our aim is to gauge the sensitivity and resolution limits at which SKA1-MID can reliably detect and morphologically 
classify FRI, FRII and star forming galaxies (SFG). We note that we do not consider any evolution in the structure of 
the radio sources with redshift, or the effects of inverse Compton scattering, but note that such
considerations would need quantifying in a more comprehensive study.
We concentrate on SKA1-MID as it provides the best possibility of investigating source morphologies when compared to 
the 
very low resolution offered by SKA1-LOW and the factor of $\sim 2-3$ poorer resolution at a similar frequency for 
SKA1-SUR. 
\subsection{Telescope Simulations}\label{sec:sims}
We use the \texttt{MAKEMS} tool to make simulated measurement sets (MS) of 8 hour tracks centred at 700~MHz, with 
fifteen 
50~MHz channels at a declination of -30 degrees. We then use the \texttt{MeqTrees} software to fill 
the MS with simulated visibilities. The noise per real and imaginary part for each visibility is calculated as
\begin{equation}
 \sigma_{\text{vis}} = \text{SEFD}/\sqrt{2\Delta\nu\Delta t}, 
\end{equation}
where $\Delta t$ is the integration time in seconds and $\Delta\nu$ is the channel width in Hertz. We use the baseline 
document's system equivalent flux density (SEFD) value of 637~Jy for band 1. Note, the noise is proportional to the 
square-root of the synthesis time, therefore since we have ``full'' uv-coverage at 8 hours, we can reasonably 
approximate a $N$ hour synthesis for $N$~$>$~8 by scaling the visibility noise as
\begin{equation}
\sigma_{\text{vis}}^N \simeq \sigma_{\text{vis}} \sqrt{8/N}.
\end{equation}
We simulated a 10 hour synthesis using the REF2 and the alternative layout W9. The noise per visibility 
for both layouts is 8.69~mJy/beam, and the root mean square (RMS) of 
the pixel noise (uniform weighting) is $5.42~\mu$Jy/beam for REF2 and $5.29~\mu$Jy/beam for W9. Note that W9 is 
slightly more sensitive compared to REF2 due to the more complete uv-coverage over the scales of interest. Primary 
beam, 
calibration and atmospheric effects are beyond the scope of this  chapter, however we note that significant progress 
has been (and continues to be) made in this area~\citep[see][]{oms2011,grobler2014,kazemi2011,kazemi2013,tasse2014}.

\subsection{Sky Models}\label{sec:sky_mods}
Starting at redshift $z=1$, we use \texttt{MeqTrees} to predict visibilities for realistic flux distributions (FRI,
FRII, or SFG galaxies) and use the \texttt{LWIMAGER} (part of the {\tt CASAREST} package) tool to make clean maps. The 
clean maps are then processed (see Section \ref{sec:classify}) to determine the morphology of the flux distributions. 
This process is repeated at incrementally higher (steps of $z=0.5$) redshifts until detection or classification is no 
longer possible. For the FRI and FRII cases, we model a flat spectrum core with two hot-spots. The core has a 
luminosity 
of 7.94$\times10^{22}$~WHz$^{-1}$sr$^{-1}$, with each hot-spot having 90\% of the core luminosity and a spectral index 
of -0.7. The core and the hot-spots are modelled as point sources and the lobes as Gaussians. For convenience we 
consider the favourable inclination angle of 45 degrees. The true size of these sources is 200~kpc. For the SFG case we 
model a a Gaussian of luminosity 1.59$\times10^{24}$~WHz$^{-1}$sr$^{-1}$ with a spectral index of -0.7, and a true size 
of 5~kpc.

\subsection{Imaging Techniques}
We make high resolution ($0.1''$) clean maps using Hogbom \texttt{CLEAN}~\citep{hogbom1974}; cleaning down to twice the 
rms pixel noise. Note that for uniform and Briggs weighting a crucial parameter is the size of the bin in the uv plane 
over which weights are ``uniformised''. By default this is determined from the full image size, but \texttt{LWIMAGER} 
allows one to uniformise the weights over bins corresponding to a user-defined FoV instead. For these simulations 
uv-bins corresponding to a FoV of 10$'$ were used.

\subsection{Morphological Classification}\label{sec:classify}
\begin{figure}[h]
 \subfloat[High SNR, $z=4$]{\includegraphics[width=0.5\textwidth]{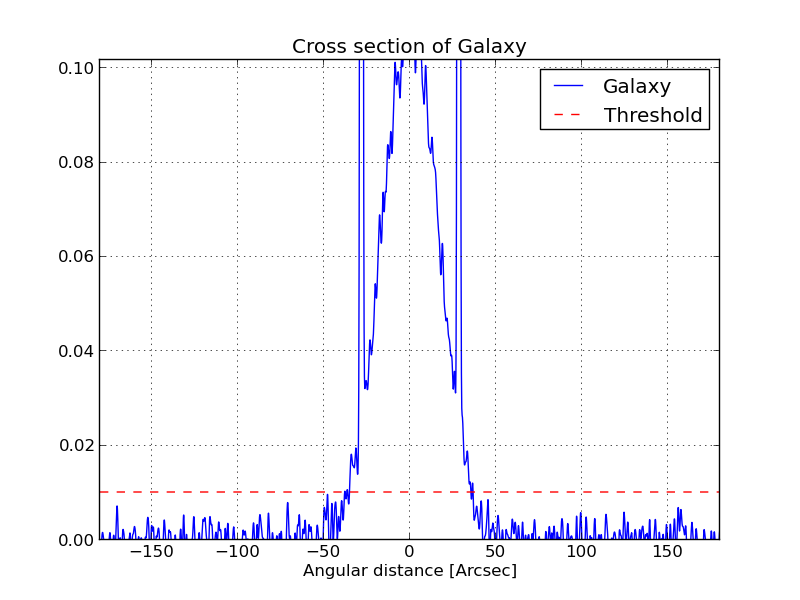}}
 \subfloat[Low SNR, $z=7$]{\includegraphics[width=0.5\textwidth]{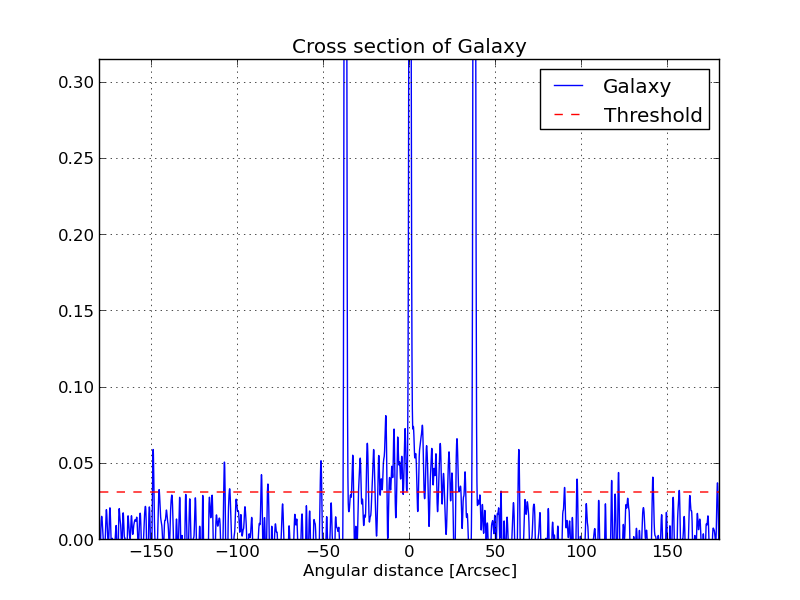}}
 \caption{Illustration of how the extent of the lobes is determined for an FR2 source. In both cases, having found the 
hot 
spots, the RMS of pixels within some box ($20\times20$ pixels in this case) along the path connecting the hot spots is 
computed. If the RMS is above some threshold (red line), then the pixels are taken to be part of the lobes. 
Finally, the edges of the lobes can be determined by keeping track of where the RMS of the pixels within this box gets 
larger than the threshold.} \label{fig:class}
\end{figure}

The classification is done in two main steps: (i) locate bright compact emission and (ii) determine the extent of 
the lobes. For the former, we use the \texttt{pyBDSM} source finding 
tool\footnote{\url{https://dl.dropboxusercontent.com/u/1948170/html/installation.html}}. If there is one 
bright compact component the galaxy is classified as a SFG, while if more than one bright component is found the 
pixel statistics along the path joining the bright components are analysed in order to determine the extent of the 
lobes -- see Figure \ref{fig:class} for an illustration. Finally, if lobes are detected, the galaxy can be classified 
as 
either FRI or FRII. This classification algorithm hinges on an accurate characterisation of the noise.

One crucial component in classifying these radio galaxies is an accurate characterisation of the jets, which is a 
major challenge at low SNR. This problem can be compounded by the inability to deconvolve the PSF at low SNR (this is 
particularly a problem for {\tt CLEAN}). However a class of algorithms based on compressive sensing (CS) has 
recently emerged to tackle deconvolution issues in the context of the SKA and its precursor/pathfinder 
facilities~\citep[see section 4.4.2 in][]{norris2013}, in particular {\tt MORESANE}~(Dabech \etal.~2014, submitted) and 
{\tt PURIFY}~\citep{carrillo2014}. Although not yet as well tested as {\tt CLEAN}, these algorithms have been shown to 
be superior to {\tt CLEAN} in detecting diffuse and compact emission, especially at low SNR\citep[][this 
volume]{ferrari2015}. In 
addition to the CS-based algorithms, a Bayesian approach to radio interferometry imaging has led to the {\tt RESOLVE} 
algorithm~\citep{jun2013}, which has also been shown to produce better results compared to {\tt CLEAN}. This progress 
in 
deconvolution algorithms will enhance the accuracy and depth of source characterisation algorithms.

\section{Simulation Results}
\begin{figure}[!ht]
 \subfloat[FRI and FRII sources]{\includegraphics[width=.5\textwidth]{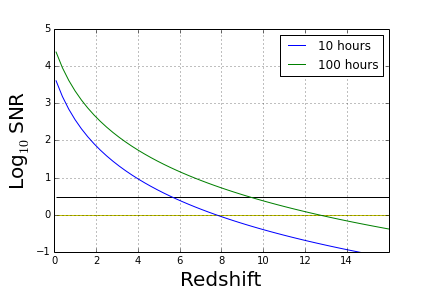}} 
 \subfloat[SFGs]{\includegraphics[width=.5\textwidth]{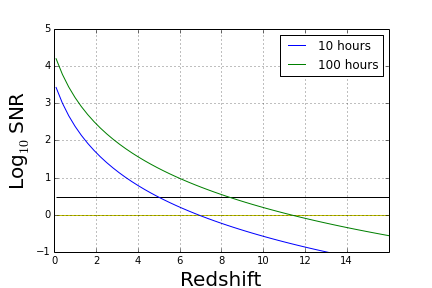}}
 \caption{The SNR (relative to the surface brightness) from a 10 (blue) and a 100 (green) hour synthesis as a function 
of redshift for FRI and FRII sources (left) and the SFGs (right). Lines of constant SNR=1 (yellow) and SNR=3 (black) 
also plotted. These plots were generated using the REF2 layout.}\label{fig:snr}
\end{figure}

\begin{figure}[!ht]
 \subfloat{\includegraphics[width=.8\textwidth]{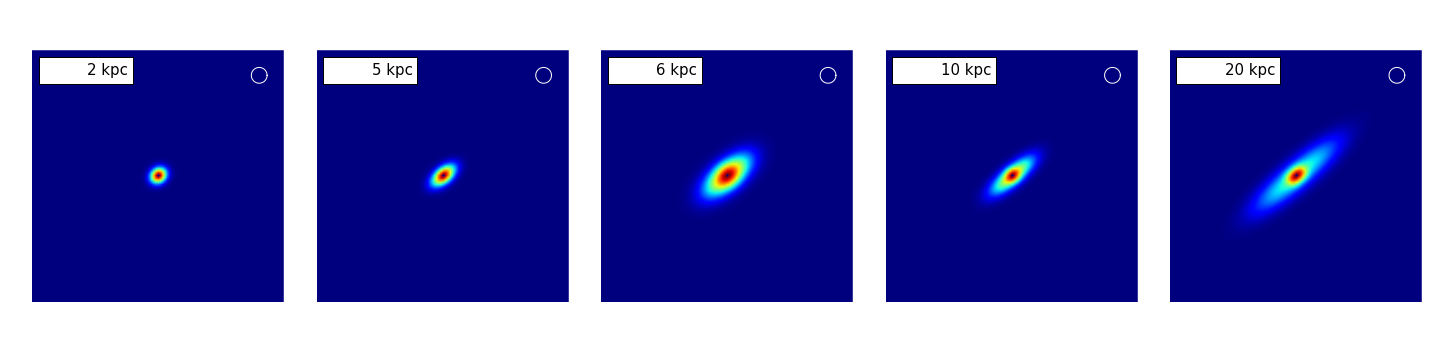}}
 \vspace{-0.7cm}
 \subfloat{\includegraphics[width=.8\textwidth]{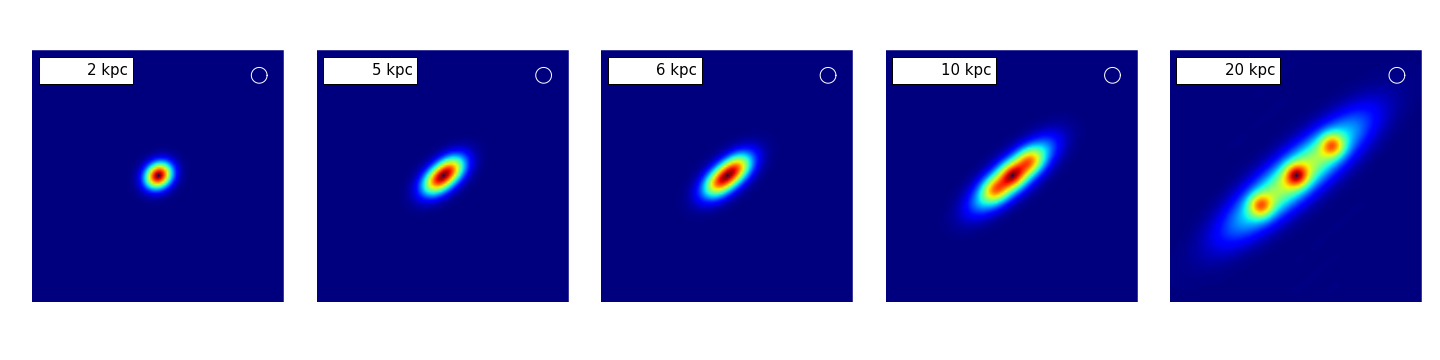}}
 \caption{Model FR1 (row 1) and FR2 (row 2) sources of different sizes as seen by the REF2 layout at $z=1$. The white 
circle on the top left corner gives an indication the PSF size (0.51$''$).}\label{fig:resolve}
\end{figure}

Before we present the simulations results, we first show in Figure \ref{fig:snr} how the SNR varies with redshift 
assuming the sky models and the observation set up described in Section \ref{sec:exp}. From these figures it is clear 
that SKA1-MID will be able to probe radio emission at very high redshifts. For FR sources we expect an SNR 
(w.r.t the lobe surface brightness for FR sources) of 3 at redshifts around 5.5, and the same SNR at a 
redshift of 5 for SFGs after 10 hours of integration and after 100 hours of integration we expect an SNR of 3 at 
redshifts of 9.5 for FR sources and 8.5 for SFGs. On the other hand, morphological characterisation will be 
limited by the ability to resolve these sources. SKA1-MID with a $\sim0.5''$ PSF (uniform weighting) at 700~MHz, will 
not be able to resolve sources of sizes less than 5~kpc (see Figure \ref{fig:resolve}).

\begin{figure}[!ht]
 \includegraphics[width=1\textwidth]{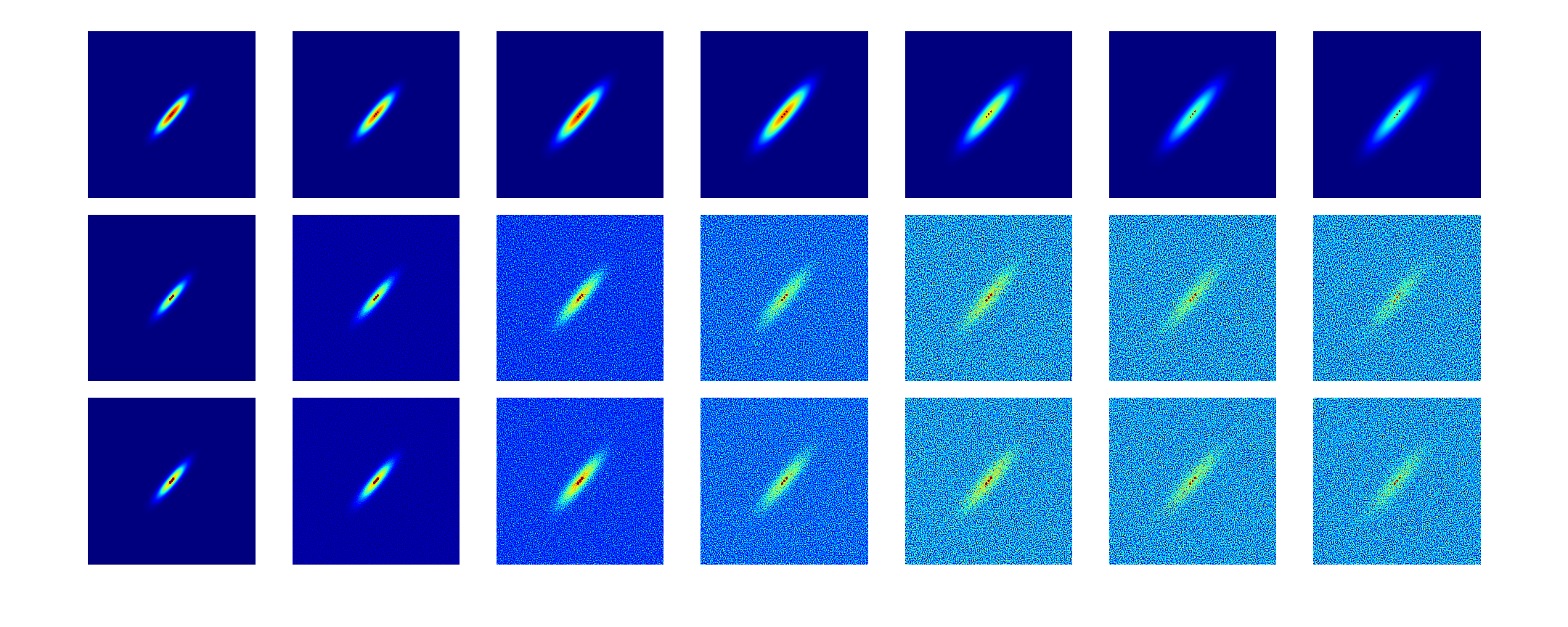}
  \caption{Model FRI galaxies (row 1) as a function of redshift \{1, 2, 3, 4, 5, 6, 7, 7.5, 8, 8.5\} as seen by the 
REF2 (row 2) and  W9 (row 3) layouts after a 10 hour synthesis.}\label{fig:fr1}
\end{figure}

\begin{figure}[!ht]
 \includegraphics[width=1\textwidth]{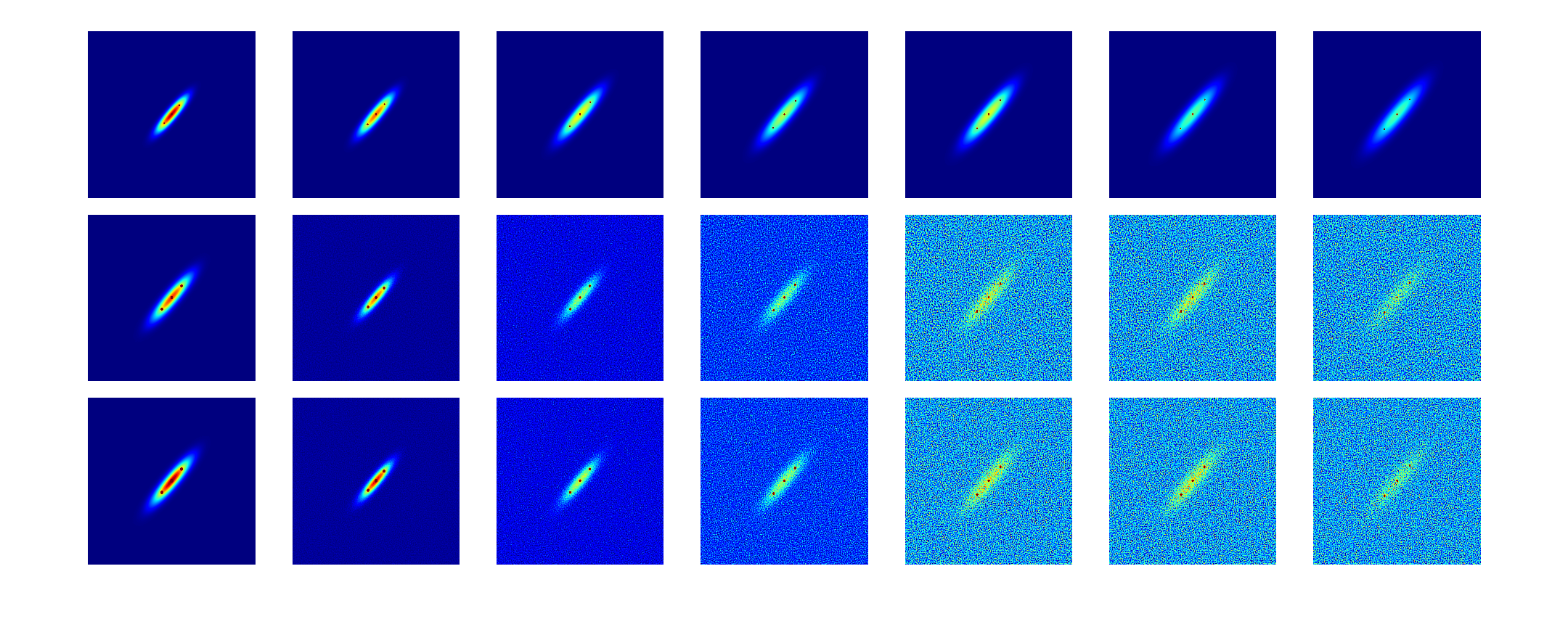}
  \caption{Model FRII galaxies (row 1) as a function of redshift \{1, 2, 3, 4, 5, 6, 7, 7.5, 8, 8.5\} as seen by the 
REF2 (row 2) and  W9 (row 3) layouts after a 10 hour synthesis.}\label{fig:fr2}
\end{figure}

\begin{figure}[!ht]
 \includegraphics[width=1\textwidth]{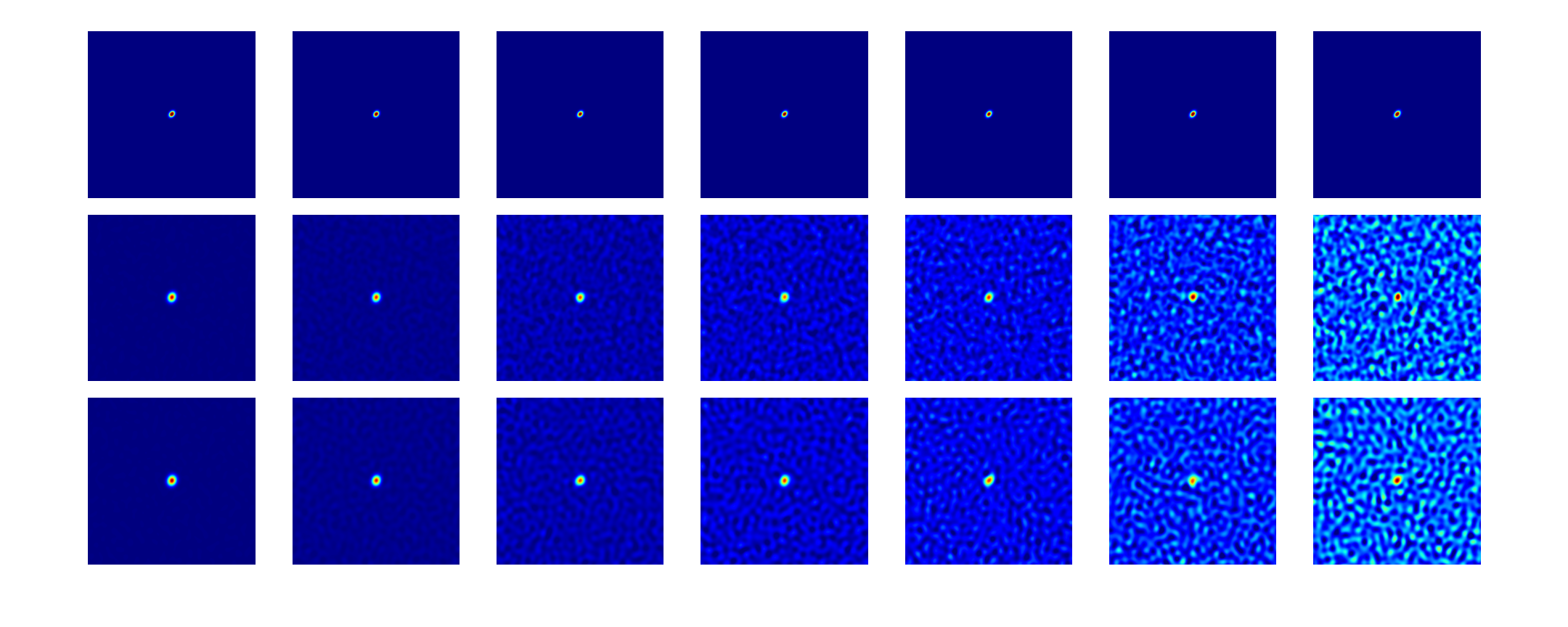}
  \caption{Model SF galaxies (row 1) as a function of redshift \{1, 2, 3, 3.5, 4, 4.5, 5\} as seen by the REF2 (row 2) 
and  W9 (row 3) layouts after a 10 hour synthesis.}\label{fig:sf}
\end{figure}

In Figures \ref{fig:fr1}-\ref{fig:sf} we show images of the galaxies under consideration as a function of redshift. For 
each galaxy type we show the sky model (row 1) and images from simulations using the REF2 and W9 layouts (rows 2 and 3 
respectively). Our morphological classification algorithm was able to detect and distinguish FR sources up to a 
redshift of 8, and up to a redshift of 5 for the SFGs after 10 hours of integration. Note that in the SFG case, we set 
a 5$\sigma$ threshold for the {\tt pyBDSM} source finder to avoid false detections, however detection and 
classification of these sources can (potentially) still be done below the 5$\sigma$ level if the false detections can 
be 
quantified using techniques such as described in \citet{serra2012}.

\begin{figure}[H]
 \subfloat[FRI and FRII sources]{\includegraphics[width=.45\textwidth]{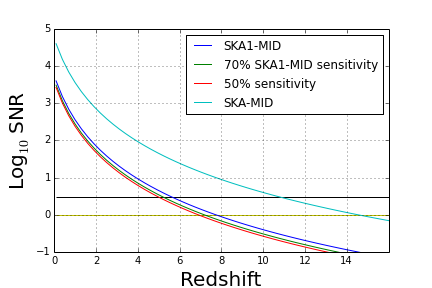}} 
 \subfloat[SFGs]{\includegraphics[width=.45\textwidth]{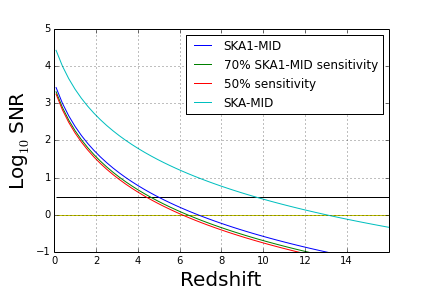}}
 \caption{The SNR (relative to the surface brightness) from a 10 hour synthesis as a function of redshift for 
FRI and FRII sources (left) and the SFGs (right). The plots are for 4 cases: (i) SKA1-MID sensitivity (blue), (ii)
70\% SKA1-MID sensitivity (green), (iii) 50\% SKA1-MID (red) and (iv) full SKA-MID sensitivity (sky blue; assuming 
SKA-MID has 10 times the sensitivity of SKA1-MID). Lines of constant SNR=1 (yellow) and SNR=3 (black) are also plotted.
}\label{fig:snr-multi}
\end{figure}

Now we consider three cases along the timeline that SKA1 may be built
out: (i) a 50\%  SKA1-MID sensitivity, (ii) a 70\% SKA1-MID in terms of sensitivity, 
and (iii) is the full SKA facility-- with 10 times the sensitivity of SKA1-MID. Figure \ref{fig:snr-multi} shows how 
the 
SNR from a 10 hour synthesis varies with redshift for these three cases.

\section{Conclusions}
Our simulation results suggest that accurate morphological classification of these high redshift radio sources will be 
feasible even at the most extreme redshifts (up to $z\sim8$ after 10 hours integration) in deep SKA1-MID images. These 
results also show that (at least for the case we have considered) an SKA1-MID layout with a maximum baseline length of 
120~km does as well as the second generation baseline layout which has a maximum baseline length of around 170~km. In 
fact, the W9 layout is slightly more sensitive to our simulated galaxies as can seen in Figures \ref{fig:fr1} and 
\ref{fig:fr2}. This would mean that for science cases where it is important to morphologically classify radio sources, 
e.g. for galaxy evolution~\citep[see][this volume]{ferramacho2014,camera2015} the proposed 120~km layout has an 
advantage, and longer baselines are not necessary.

Morphological classification will be limited by resolution and not sensitivity. SKA1-MID will be able to resolve 
radio sources down to scales of $\sim5$~kpc at cosmological distances. Morphological classification at smaller scales 
will require higher resolution. This is only available by going to higher frequencies (though impractical in a survey 
scenario due to the reduced field of view), or with substantially longer baselines, such as those provided by the full 
SKA dish array with stations in SKA partner countries.

It is also worth noting that model fitting techniques~\citep{vidal2014,white1997,reid2006} have the capability 
to ``super-resolve'' multiple components within the PSF main lobe. Given the high sensitivity of SKA1-MID, the 
prospects of morphologically separating star formation and AGN activity are very good since the ability to infer the 
morphology of sub-resolution sources scales as $\sqrt{SNR}$~\citep{vidal2014}. In particular, with SKA1-MID achieving 
an SNR of 10 to 100 in the range z=$1\sim4$, this translates into the ability to resolve features of roughly one third 
to a 10th of the PSF size via model fitting, thus making morphological classification of kpc-scale sources possible.

A more general algorithm (compared to the one presented here) will be required to do this classification in deep field 
images with multiple sources. Such algorithms can be further improved by considering spectral and polarization 
information.

\acknowledgments
S. Makhathini acknowledges financial support from the National Research Foundation of South
Africa. O. Smirnov's research is supported by the South African Research Chairs Initiative of the 
Department of Science and Technology and National Research Foundation.

\bibliographystyle{apj_short_etal}
\bibliography{makhathini}
\end{document}